\documentclass[proof]{pasj00}

\begin{document}
\SetRunningHead{Author(s) in page-head}{Running Head}
\Received{2011/01/19}
\Accepted{}

\title{Particle-Particle Particle-Tree: A Direct-Tree Hybrid Scheme for
Collisional $N$-Body Simulations}

\author{Shoichi \textsc{Oshino}\altaffilmark{1,2},
Yoko \textsc{Funato}\altaffilmark{4},
and Junichiro \textsc{Makino}\altaffilmark{1,2,3}}

\altaffiltext{1}{Department of Astronomical Science, The Graduate
University for Advanced Studies (SOKENDAI), 2-21-1 Osawa, Mitaka, Tokyo
181-8588}

\altaffiltext{2}{Division of Theoretical Astronomy, National
Astronomical Observatory of Japan, 2-21-1 Osawa, Mitaka, Tokyo
181-8588}
\email{oshino@cfca.jp}
\altaffiltext{3}{Center for Computational Astrophysics, National
Astronomical Observatory of Japan, 2-21-1 Osawa, Mitaka, Tokyo
181-8588}
 \email{makino@cfca.jp}
\altaffiltext{4}{Department of General System Studies, Graduate School
 of Arts and Sciences, The University of Tokyo, 3-8-1 Komaba, Meguro,
 Tokyo 153-8902}
\email{funato@artcompsci.org}


%


\maketitle

\begin{abstract}
In this paper, we present a new hybrid algorithm for the time
 integration of collisional $N$-body systems. In this algorithm,
 gravitational force between two particles is divided into short-range
 and long-range terms, using a distance-dependent cutoff function. The
 long-range interaction is calculated using the tree algorithm and
 integrated with the constant-timestep leapfrog integrator. The
 short-range term is calculated directly and integrated with the
 high-order Hermite scheme. We can reduce the calculation cost per
 orbital period from $O(N^2)$ to $O(N \log N)$, without significantly
 increasing the long-term integration error. The results of our test
 simulations show that close encounters are integrated
 accurately. Long-term errors of the total energy shows random-walk
 behaviour, because it is dominated by the error caused by tree
 approximation.
\end{abstract}

 \section{Introduction} 
For the time integration of collisional $N$-body systems such as star
clusters and systems of planetesimals, the combination of direct
summation for force calculation and the individual timestep algorithm
has been the standard method for nearly half century (\cite{a63},
\yearcite{a03}). For galactic dynamics and cosmology, fast and
approximate methods for force calculation such as the particle-mesh
scheme (\cite{he81}), the $\rm P^3M$ scheme (\cite{he81}), the tree
method (\cite{bh86}) and combinations of PM and tree (\cite{x95},
\cite{b02}, \cite{dkph04}, \cite{s05}, \cite{yf05}, \cite{ifm09}) are
used. It is not impossible to combine individual timestep algorithm and
fast and approximate force calculation. For example, \citet{ma93}
developed a high-order integrator using individual timestep combined
with the tree algorithm. However, it was difficult to achieve good
performance on distributed-memory parallel computers for such
scheme. \citet{fifm07} introduced a hybrid of tree and individual
timestep algorithm, which is designed to handle the evolution of star
clusters embedded in the parent galaxy. In their BRIDGE scheme, both the
parent galaxy and the star cluster are expressed as $N$-body
systems. The interaction between particles in the star cluster are
calculated directly and integrated with the individual timestep scheme,
while interactions between particles in the parent galaxy and that
between particles in parent galaxy and particles in star clusters are
calculated with the tree algorithm and integrated with the leapfrog
scheme with shared and constant timesteps.

The BRIDGE scheme is based on the idea of splitting the Hamiltonian of
an $N$-body system to multiple components. The time integration of
tree part is symplectic and does not generate any secular error. The
direct integration of the internal motion of star cluster is not
symplectic, but treated with high accuracy using high-order
integrators. The obvious limitation of the BRIDGE scheme is that it can
handle the close encounters of particles in star clusters only. If we
want to allow close encounter between particles in the parent galaxy, it
goes back to the usual direct summation scheme.

For the time integration of planetary systems, the mixed variable
symplectic (hereafter MVS) scheme (\cite{kyn91}, \cite{wh91}) has become
the standard method. For the time integration of almost stable orbits of
planets, the MVS is well suited. However, if we want to handle
protoplanets or planetesimals, we need to handle their close encounters
and collisions. The MVS scheme, however, cannot handle them since it
requires that the timestep is kept constant (\cite{cs93}). In order to
handle close encounters, several modifications of MVS method have been
proposed. One is the hybrid method by \citet{c99}. It handles close
encounters with high accuracy by splitting the Hamiltonian to that of
close interaction and distant interaction as in the case of $\rm P^3M$
method. It can be used for planetary accretion problems. However, it
relies on the direct calculation and its calculation cost scales as
$O(N^2)$. Very recently, \citet{mqe08} applied GPGPU to the hybrid
method. In addition, they used the Hermite scheme (\cite{ma92}) for the
integration of the short range force.

\citet{bv03} and \citet{bsvc07} combined the hybrid method and the tree
algorithm.  Therefore their method can handle a large number of particles and
close encounters. The scheme described in \citet{bv03} uses the leapfrog
integrator to handle the Hamiltonian for the short-range
interaction. Therefore it had to use very small timesteps. The scheme
described in \citet{bsvc07} seems to be improved, but no details of the
scheme was given.

In this paper, we describe a new time integration algorithm which
combines the strong points of these schemes. This scheme is combination
of the BRIDGE scheme (\cite{fifm07}) and a hybrid symplectic integrator
method (\cite{c99}). We call this scheme as Particle-Particle
Particle-Tree (hereafter PPPT). In section 2 we overview previous
numerical methods. These methods are based on the MVS method. Therefore,
we first introduce the MVS method and then overview extension of MVS
method. In section 3 we describe our new time integration algorithm,
PPPT. In section 4, we show the result of test simulations. In this
paper, we discuss only the simulations of planet formation
process. However, in future works, we apply PPPT to other collisional
systems. Summary and discussions are given in section 5.

\section{The Numerical Method} 

\subsection{The symplectic integrator} 

The Hamiltonian of an $N$-body system is given by
\begin{equation}
 H=\sum_i^N\frac{p_i^2}{2m_i}-\sum_{i=1}^{N-1}\sum_{j=i+1}^N\frac{Gm_im_j}{r_{ij}},\label{H_n}
\end{equation}
where $p_i$ is the momentum of particle $i$, $r_{ij}$ is the distance
between particles $i$ and $j$, $m_i$ is the mass of particle $i$ and $G$
is the gravitational constant. The Hamilton's equation of motion is
\begin{equation}
 \frac{{\rm d}f}{{\rm d}t}=\{f,H\},\label{em_n}
\end{equation}
where $\{f,H\}$ is the Poisson bracket and $f$ is a canonical variable.
We define the differential operator as $Df\equiv \{f,H\}$. The general
solution of equation (\ref{em_n}) at time $t+\Delta t$ is formally written as
\begin{equation}
 f(t+\Delta t)={\rm e}^{\Delta t D}f(t). \label{te_of_f}
\end{equation}
Equation (\ref{te_of_f}) cannot be solved analytically. We can obtain
the approximate solution by dividing the Hamiltonian into multiple
parts that can be analytically solved.

In the symplectic integrator, the Hamiltonian is divided into two parts
as
\begin{eqnarray}
 H &=& H_A+H_B,\\
 H_A &=& -\sum_{i=1}^{N-1}\sum_{j=i+1}^N\frac{Gm_im_j}{r_{ij}},\\
 H_B &=& \sum_i^N\frac{p_i^2}{2m_i},
\end{eqnarray}
where $H_A$ is the potential energy and $H_B$ is the kinetic energy, and
each of which can be solved. The time evolution of $f$ is given by
\begin{equation}
 f(t+\Delta t)={\rm e}^{\Delta t D}f(t)={\rm e}^{\Delta t (A+B)}f(t),\label{te_AB}
\end{equation}
where $A\equiv \{,H_A\}$ and $B\equiv \{,H_B\}$ are operators. The
exponential in equation (\ref{te_AB}) can be approximated as
\begin{equation}
 {\rm e}^{\Delta t(A+B)}=\prod_{i=1}^{k}{\rm e}^{a_i \Delta t A}{\rm
  e}^{b_i \Delta t B} + O(\Delta t^{n+1}), \label{ex}
\end{equation}
where $(a_1,a_2,...,a_k)$ and $(b_1,b_2,...,b_k)$ are real numbers, $k$
is the number of stages, $n$ is the order of approximation. The first
term of the right hand side of equation (\ref{ex}) is an order-$n$
approximation of the left hand side. The first-order symplectic
integrator is given by
\begin{equation}
  f(t+\Delta t)={\rm e}^{\Delta t A}{\rm e}^{\Delta t B}f(t) + O(\Delta t^{2}),
\end{equation}
and the second-order symplectic integrator, which is the leapfrog
integrator, is given by
\begin{equation}
 f(t+\Delta t)={\rm e}^{\Delta t A/2}{\rm e}^{\Delta t B}{\rm e}^{\Delta t A/2}f(t) + O(\Delta t^{3}).\label{em_l}
\end{equation}
The symplectic integrator has the advantage that there is no long-term
energy error for the time integration of periodic systems. In the case
of a near-Kepler potential, both the semi-major axis and the
eccentricity are conserved. One disadvantage of the symplectic method is
that high-order schemes are expensive and have rather large local error
coefficients. Another disadvantage is that it requires a constant
timestep. If we change the timestep following usual recipes for
Runge-Kutta-Fehlberg-type schemes (\cite{f68}), the energy is no longer
conserved (\cite{sg92}, \cite{cs93}). There are a number of proposals
for methods to combine the variable timesteps and the good nature of
the symplectic schemes. Most of them are based on the idea of splitting
the potential to fast-varying and second-varying terms and applying
different timesteps.  For example, \citet{sb94} proposed a method that
splits the gravitational force into many components, each of which has
finite effective range in the distance. By assigning different timesteps
to different components, they effectively realized a variable timestep
integration.

The methods discussed below are all based on the idea of changing the
way to split the Hamiltonian.

\subsection{The Mixed Variable Symplectic Method}

\citet{kyn91} and then \citet{wh91} introduced the MVS method for
 planetary systems. In the symplectic scheme, we split the Hamiltonian
 to kinetic and potential energy, so that we have analytical solutions
 for both parts. This division is not unique. As far as each splitted
 Hamiltonian has an analytic solution, any division can be used. The
 idea of MVS is to split the Hamiltonian to the Keplarian term $H_{Kep}$
 and interaction term $H_{Int}$. The time evolution is given by
\begin{eqnarray}
 H &=& H_{Kep}+H_{Int},\\
 f(t+\Delta t) &=& {\rm e}^{\Delta t K/2}{\rm e}^{\Delta t I}{\rm e}^{\Delta t K/2}f(t),
\end{eqnarray}
where $K$ is the operator defined as $Kf\equiv \{f,H_{Kep}\}$ and $I$ is
the operator defined as $If\equiv \{f,H_{Int}\}$. The MVS method
integrates $H_{Int}$ with the leapfrog integrator and $H_{Kep}$ by using
analytic solution of the Kepler orbit. Thus the MVS method is expressed
as follows:
\begin{enumerate}
 \item Calculate accelerations due to gravitational interactions
       between planets at time $t$ and give a velocity kick.
 \item Update analytically positions and velocities from $t$ to $t+\Delta
       t$ by using the Solar gravity.
 \item Calculate accelerations due to gravitational interactions
       between planets at time $t+\Delta t$ and give a velocity kick.
\end{enumerate}
The advantage of MVS is that only interactions between planets are
integrated numerically. The Solar gravity is analytically integrated and
is accurate up to the round-off error.

\subsection{The BRIDGE code}

The BRIDGE code was introduced by \citet{fifm07}. It was designed for
the time integration of star clusters embedded in parent galaxies. This
scheme is a combination of the direct and a tree schemes, using the idea
similar to that of the MVS scheme. The internal interactions of stars in
star clusters are integrated by the Hermite integrator with direct
summation, while other parts are integrated by the leapfrog integrator
and tree scheme.

The BRIDGE scheme divides the Hamiltonian as
\begin{eqnarray}
 H &=& H_{\alpha}+H_{\beta},\\
 H_{\alpha} &=& -\sum_{i<j}^{N_{G}}
  \frac{Gm_{G,i}m_{G,j}}{r_{GG,ij}}-\sum_{i=1}^{N_{G}} \sum_{j=1}^{N_{SC}}
  \frac{Gm_{G,i}m_{C,j}}{r_{GC,ij}},\\
 H_{\beta} &=& \sum_{i=1}^{N_{G}} \frac{p^2_{G,i}}{2m_{G,i}}+\sum_{i=1}^{N_{SC}}
  \frac{p^2_{C,i}}{2m_{C,i}}-\sum_{i<j}^{N_{SC}} \frac{Gm_{C,i}m_{C,j}}{r_{CC,ij}},
\end{eqnarray}
where $N_G$ and $N_{SC}$ are the number of particles in the parent
galaxy and that in the star cluster, $m_{G,i}$ and $p_{G,i}$ are the
mass and the momentum of particle $i$ in the galaxy, $m_{C,i}$ and
$p_{C,i}$ are those of particle $i$ in the star cluster, and $r_{GG},
r_{GC}, r_{CC}$ are distances between two galaxy particles, one galaxy
and one star cluster particle, and two star cluster particles,
respectively. We can express the time evolution from $t$ to $t+\Delta t$
as
\begin{eqnarray}
 f(t+\Delta t)=e^{\frac{1}{2}\Delta t\alpha}e^{\Delta
  t\beta}e^{\frac{1}{2}\Delta t\alpha}f(t),
\end{eqnarray}
where $\alpha$ is the operator defined as $\alpha f \equiv
\{f,H_{\alpha}\}$ and $\beta$ is the operator defined as $\beta f \equiv
\{f,H_{\beta}\}$.

The BRIDGE code uses the leapfrog scheme for $H_{\alpha}$, and the
fourth-order Hermite scheme for $H_{\beta}$. Thus the integration
procedure during a tree timestep $\Delta t$ is done in the following way:
\begin{enumerate}
 \item Make a tree at time $t$ and calculate accelerations from all particles
       on galaxy particles, and from galaxy particles on star-cluster
       particles.
 \item Give a velocity kick for star cluster particles, and update the
       velocities of galaxy particles.
 \item Integrate positions and velocities of star cluster particles from
       $t$ to $t+\Delta t$ by using the Hermite scheme with the
       individual timestep, and positions of galaxy particles by making
       them drift with the constant velocities.
 \item Make a new tree at $t+\Delta t$ and calculate accelerations from
       all particles to galaxy particles, and from galaxy particles to
       star cluster particles.
 \item Give a velocity kick for star cluster particles and update
       velocities for galaxy particles.
\end{enumerate}
In this scheme, $H_{\alpha}$ is integrated with the symplectic leapfrog
scheme, while $H_{\beta}$ is integrated with non-symplectic Hermite
scheme. It combines the fast tree code for the orbital motion
of particles in the galaxy and high-accuracy Hermite scheme for the
internal orbital motion of particles in the star cluster without any
additional approximation. Thus, it is the first scheme with which we can
follow the orbital and internal evolution of a star cluster embedded in
a galaxy in a fully self-consistent way.

\subsection{The MERCURY code}

The MERCURY code (\cite{c99}) splits the Hamiltonian into three parts as
\begin{eqnarray}
 H &=& H_{Kep}+H_{Int}+H_{Sun}, \label{eq:H}\\
 H_{Kep} &=& \sum_{i=1}^{N}\left(\frac{p^2_i}{2m_i}-\frac{Gm_im_{\odot}}{r_i}\right)
  -\sum_{i<j}^{N}\frac{Gm_im_j}{r_{ij}}W(r_{ij}),\\
 H_{Int} &=& -\sum_{i<j}^{N}\frac{Gm_im_j}{r_{ij}}(1-W(r_{ij})), \label{eq:Hint}\\
 H_{Sun} &=& \sum_{i=1}^N\frac{p^2_i}{2m_{\odot}},
\end{eqnarray}
where $H_{Int}$ is the potential energy of gravitational interactions
between particles except for those undergoing close encounters,
$H_{Kep}$ is the kinetic energy of particles plus potential energy of
particles undergoing close encounters, $H_{Sun}$ is the kinetic energy
of the Sun, and $W(r_{ij})$ is the changeover function. The modification
from MVS is that the potential energy of nearby particles is separated
from $H_{Int}$ and moved to $H_{Kep}$. The time evolution is described
as
\begin{eqnarray}
 f(t+\Delta t)=e^{\frac{1}{2}\Delta t I}e^{\frac{1}{2}\Delta t S}e^{\Delta
  t K}e^{\frac{1}{2}\Delta t S}e^{\frac{1}{2}\Delta t I}f(t).
\end{eqnarray}
where $I$ is the operator defined as $I f \equiv \{f,H_{I}\}$, $S$ is
the operator defined as $S f \equiv \{f,H_{S}\}$ and $K$ is the operator
defined as $K f \equiv \{f,H_{K}\}$. This scheme is sometimes called the
hybrid scheme. In the actual code (MERCURY), the force from
gravitational interactions is split instead of the Hamiltonian for the
ease of programing. We can write the pairwise force $ F(r_{ij})$ as
\begin{eqnarray}
 F(r_{ij}) &=& F_{close}(r_{ij})+F_{dist}(r_{ij}), \label{eq:F}\\
 F_{close}(r_{ij}) &=& F(r_{ij})K(r_{ij}),\\
 F_{dist}(r_{ij}) &=& F(r_{ij})[1-K(r_{ij})],\label{eq:Fdist}
\end{eqnarray}
where $F_{close}(r_{ij})$ is the force for close encounter and
$F_{dist}(r_{ij})$ is the remaining force. The relation between $W$ and
$K$ is given by
\begin{equation}
 W(r) = r \int_{r}^{\infty} \frac{K(x)}{x^2}dx
\end{equation}
In the hybrid scheme, $H_{Kep}$ dose not have an analytical solution if
the potential of close encounter is not zero. Therefore, $Kf$ is
integrated numerically using the Bulirsch-Stoer method (\cite{bs64} and
\cite{sb80}). The one-step integration of the hybrid scheme is done in
the following way:
\begin{enumerate}
 \item Apply the velocity kick, $e^{\Delta \frac{1}{2}t I}$, due to
       distant interaction $H_{Int}$.
 \item Apply the position drift $e^{\Delta \frac{1}{2}t S}$, due to $H_{Sun}$.
 \item Integrate orbits of particles which are under close
       encounters using BS method, and update positions and velocities
       of the rest of particles using the Kepler orbit.
 \item Give the position drift by $H_{Sun}$ with stepsize $\Delta t/2$.
 \item Give the velocity kick by $H_{Int}$ with stepsize $\Delta t/2$.
\end{enumerate}

The hybrid method is used for long-term integrations of outer solar
system, restricted three-body problems, and planetary embryos. It has a
good performance for accuracy and speed for small-$N$ systems, but for
system with a large number of particles it becomes expensive.

\subsection{The DAEDALUS code}

\citet{bsvc07} presented a new mixed-variables symplectic tree code for
planetesimal dynamics, DAEDALUS. This code is an improved version of the modified
tree code described in \citet{bv03}. \citet{bv03} developed the tree
code (\cite{bh86}) with two-level timesteps for planetesimal
systems. This tree code usually integrates all particles with a constant
timestep using the leapfrog integrator. If close encounters occur, it
integrates only particles which undergo close encounters with much
smaller timesteps. When integrating close encounters, timesteps are
determined using equation (2) in \citet{bv03}, and the tree is
constructed at each steps.

The DAEDALUS integrator is combination of the tree code and a hybrid
symplectic integrator method. It splits the Hamiltonian following the
description of \citet{c99}. Therefore the DAEDALUS integrator integrates
$H_{Int}$ using tree method and integrates $H_{Kep}$ analytically where
there are no close encounters. If close encounters occur, it integrates
$H_{Kep}$ numerically using the Bulirsch-Stoer method. The one-step
integration of the DAEDALUS integrator is done in the following way:
\begin{enumerate}
 \item Make a tree and calculate accelerations from $H_{Int}$ then give
       a velocity kick.
 \item Give a position drift by $H_{Sun}$ with stepsize $\Delta t/2$.
 \item Calculate positions and velocities for particles which are in
       close encounters, and update positions and velocities with the
       Kepler orbit for particles which are not in close encounters.
 \item Apply the position drift $e^{\Delta \frac{1}{2}t S}$, due to $H_{Sun}$.
 \item Make a new tree at next step and calculate accelerations from
       $H_{Int}$ then give a velocity kick.
\end{enumerate}

\begin{table}
 \begin{tabular}{|c||c|c|c|c|}
  \hline
  & Tree method & distant-based criterion & variable timestep &
  individual timestep  \\
  \hline
  \hline
  MVS        & - & - & - & -   \\
  MERCURY    & - & $\bigcirc$ & $\bigcirc$ & - \\
  BRIDGE     & $\bigcirc$ & - & $\bigcirc$ & $\bigcirc$ \\
  DAEDALUS   & $\bigcirc$ & $\bigcirc$ & $\bigcirc$ & - \\
  This paper & $\bigcirc$ & $\bigcirc$ & $\bigcirc$ & $\bigcirc$ \\
  \hline
 \end{tabular}
\caption{Characteristics of each method.}
\end{table}

The DAEDALUS integrator uses the variable, but shared, timestep for
close encounters. Therefore, when the number of particles in close
encounters is large, the calculation cost can become high. If we use an
individual timestep, we can decrease the calculation cost significantly.
In Table 1, we summarize the characteristics of the previous and present
algorithms. Table 1 shows that there were no algorithms which has all of
the listed desirable properties except for the one described in this
paper. In this paper, we present a new algorithm which uses the shared
timestep for distant interactions and the individual timestep for close
interactions. Furthermore, the force due to distant interactions are
calculated by using tree method with a changeover function.

\section{Particle-Particle Particle-Tree (PPPT) scheme} 

In this section we describe our new scheme for collisional $N$-body
systems. We use the fourth-order Hermite method for near-neighbour
forces and the tree method for distant forces, and we use the hybrid
method to split the gravitational force by using a changeover
function. In our scheme, we split the gravitational force between two
particles as
\begin{eqnarray}
 F(r_{ij}) &=& F_{Hard}(r_{ij})+F_{Soft}(r_{ij}),\\
 F_{Hard}(r_{ij}) &=& F(r_{ij})K(r_{ij}),\label{eq:fk}\\
 F_{Soft}(r_{ij}) &=& F(r_{ij})[1-K(r_{ij})],
\end{eqnarray}
where $F_{Hard}(r_{ij})$ is the force from nearby particles and
$F_{Soft}(r_{ij})$ is the force from distant particles. These formulae
are the same as equations (\ref{eq:F})-(\ref{eq:Fdist}). We can write
the Hamiltonian as
\begin{eqnarray}
 H &=& H_{Hard}+H_{Soft},\\
 H_{Hard} &=& \sum_{i=1}^{N}\left(\frac{p^2_i}{2m_i}-\frac{Gm_im_{\odot}}{r_i}\right)
-\sum_{i<j}^{N}\frac{Gm_im_j}{r_{ij}}W(r_{ij}),\\
 H_{Soft} &=& \sum_{i<j}^{N}\frac{Gm_im_j}{r_{ij}}(1-W(r_{ij})),
\end{eqnarray}
where $H_{Hard}$ contains the kinetic energy of all particles and the
potential energy of near-neighbours, and $H_{Soft}$ contains the
potential energy of all other pairs of particles. We treat the solar
gravity as the force caused by a fixed potential in this paper. So we do not have
$H_{Sun}$ here. We can express the time evolution from $t$ to $t+\Delta
t$ as
\begin{eqnarray}
 f(t+\Delta t)=e^{\frac{1}{2}\Delta t S}e^{\Delta
  t H}e^{\frac{1}{2}\Delta t S}f(t).
\end{eqnarray}

\begin{figure}
  \begin{center}
   \FigureFile(80mm,80mm){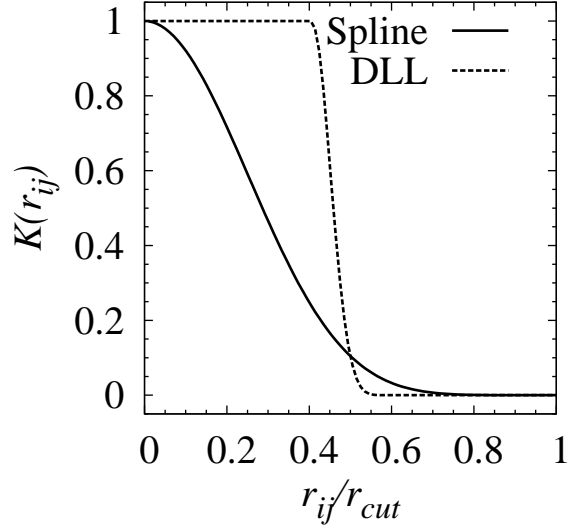}
  \end{center}
  \caption{The changeover functions. The horizontal axis is the
  distance between $i$ and $j$ particles normalized by the cutoff
  radius, and $K(r_{ij})$ is the changeover function. The solid and
  dashed curve, show the fourth order spline function and the DLL
  function. This function uses $r_1/r_{cut}=0.4$ and
  $r_2/r_{cut}=0.6$.}\label{fig:change}
 \end{figure}

The changeover function $K$ splits the gravitational force between
particles to contributions of close encounters and others. Thus,
$H_{Hard}$ changes rapidly and $H_{Soft}$ changes slowly. In this paper,
we use two types of changeover functions. One is the fourth order spline
function which introduced by \citet{asio86} given as
\begin{equation}
 K(r_{ij}) = \left(\frac{\sin X}{X} \right)^5,\label{eq:spline}
\end{equation}
where $X=\pi r_{ij}/r_{cut}$ and $r_{cut}$ is a scaling radius. Note that $K(r_{ij})$ becomes zero where
$r_{ij}\geq r_{cut}$ (see fig \ref{fig:change}). The other was first introduced by \citet{ld00}. It is
given by
\begin{equation}
K(r_{ij})=  \left\{
\begin{array}{@{\,}ll}
 1 & {\rm if} \ Y \geq 1, \\
 10Y^6 -15Y^8 +6Y^{10}
  & {\rm if} \ 0<Y<1, \\
 0 & {\rm if} \ Y \le 0,
\end{array}
\right.
\end{equation}
where $Y=\frac{r_2-r_{ij}}{r_2-r_1}$. Hereafter we call it the DLL
function. In figure \ref{fig:change}, $r_1/r_{cut}=0.4$ and
$r_2/r_{cut}=0.6$. We call $r_2$ the cutoff radius of the DLL function.

In this paper we regard the gravitational field of the Sun as an
external potential. This treatment is okay for the study of planet
formation process of earth-type planets, because planetesimals do not
perturb the Sun strongly. The total mass of planetesimals is much
smaller than the mass of the Sun and planetesimals are distributed
almost uniformly around the Sun.

We integrate $H_{Soft}$ using the leapfrog and the tree method with a
constant timestep $\Delta t$. We integrate $H_{Hard}$ using the
fourth-order Hermite method with the block timestep. For the timestep
criterion, we used a slightly modified version of the ``standard''
criterion (\cite{ma92} and \cite{a03}). The standard criterion is given by
\begin{eqnarray}
 \Delta
  t_i&=&\eta\sqrt{\frac{|a_i^2||a_i^{(2)}|+|\dot{a_i}|^2}{|\dot{a_i}||a_i^{(3)}|+|a_i^{(2)}|^2}}, \label{eq:standard}
\end{eqnarray}
where $\eta$ is the accuracy parameter. The new timestep criterion we used is
given as
\begin{eqnarray}
 \Delta
  t_i&=&\eta\sqrt{\frac{\sqrt{a_i^2+a_0^2}|a_i^{(2)}|+|\dot{a_i}|^2}{|\dot{a_i}||a_i^{(3)}|+|a_i^{(2)}|^2}},\\ \label{eq:timestep}
 a_0 &=& \alpha^2 \frac{Gm_i}{r_{H}^2}.\label{eq:alpha}
\end{eqnarray}
Here $a_0$ is a constant introduced to prevent the timestep from
becoming unnecessarily small when $|a_i|$ is small, and $\alpha$ is a
parameter which controls the size of the timestep. Here, $a_i$ is the
acceleration due to $F_{Hard}$, and $\dot{a_i}, a_i^{(2)}$ and
$a_i^{(3)}$ are its first, second and third time derivatives,
respectively, and $m_i$ is the mass of particle. With equation
(\ref{eq:standard}), the timestep becomes unnecessarily small if there
is just one particle inside the radius $r_{cut}$ of one particle and
$r_{ij} \simeq r_{cut}$. To illustrate this problem, consider the case
in which one particle moves away radially with a constant velocity $v$.
High order derivatives of the force from this particle is given by
\begin{eqnarray}
 a &=& FK,\\ \label{eq:a}
 \dot{a} &=& (F'K+FK')v,\\
 a^{(2)} &=& (F''K+2F'K'+FK'')v^2,\\
 a^{(3)} &=& (F'''K+3F''K'+3F'K''+FK''')v^3 \label{eq:a3}
\end{eqnarray}
for the equation (\ref{eq:standard}), where $F$ is the gravitational
force from a particle and $K$ is the changeover function and $F'(x)={\rm
d}F/{\rm d}x$. In the case of
$r_{ij} \lesssim r_{cut}$, we can expand $K$ around $r_{cut}$. Without
the loss of generality, we can assume that $r_{cut}=1$ and
$\mathrm{d} r / \mathrm{d} t =v=1$. Then we have
\begin{eqnarray}
 K &=& -Z^5 + O(Z^6), \\ \label{eq:k}
 K' &=& -5Z^4 + O(Z^5), \\
 K'' &=& -20Z^3 + O(Z^4), \\
 K''' &=& -60Z^2 + O(Z^3), \label{eq:k3}
\end{eqnarray}
where $Z \equiv \left(\frac{r_{ij}-r_{cut}}{r_{cut}} \right)$.
By substituting equations (\ref{eq:k})-(\ref{eq:k3}) into equations
(\ref{eq:a})-(\ref{eq:a3}), and omitting time derivatives of
gravitational force, we obtain
\begin{eqnarray}
 a &=& -Z^5F + O(Z^6),\\\label{eq:expand_a}
 \dot{a} &=& -5Z^4F + O(Z^5),\\
 a^{(2)} &=& -20Z^3F + O(Z^4),\\
 a^{(3)} &=& -60Z^2F + O(Z^3).\label{eq:expand_a3}
\end{eqnarray}
Because time derivatives of gravitational force are individual to $Z$.
Equation (\ref{eq:standard}) becomes
\begin{eqnarray}
 \Delta t &=& \eta \sqrt{ \frac{45Z^8F^2 + O(Z^9)}{700Z^6F^2 + O(Z^7)} },\\
  &=& \eta Z \sqrt{ \frac{9 + O(Z)}{140 + O(Z)} },\\
  &=& \eta Z(1260 + O(Z)). \label{eq:ex_timestep}
\end{eqnarray}
Equation (\ref{eq:ex_timestep}) shows that the timestep approaches to zero
as $r_{ij}$ approaches to $r_{cut}$. This behaviour is clearly
undesirable, since there is no need to reduce the timestep for the
neighbour force, when the neighbour force itself is small. The
reason why criterion (\ref{eq:standard}) gives zero stepsize is, as we
can see from equations (\ref{eq:expand_a})-(\ref{eq:expand_a3}), $|a|$
approaches to zero faster than its high order derivatives. However, this
is due to the cutoff by the changeover function, and the actual physical
acceleration by the particle just inside radius $r_{cut}$ is of the order
$\frac{m_j}{r_{cut}^2}$. In order to avoid this unnecessarily small
timestep, we introduce $a_0$ in equation (\ref{eq:timestep}). By doing
so, we make criterion (\ref{eq:alpha}) to give the timestep which is
accurate relative to the absolute strength of the force itself, before
the changeover function is applied. This timestep criterion does not use
the gravitational force from the Sun. In other words, the timestep is
determined purely by the forces from nearby particles. If we included
the Solar gravity, the original criterion could lead to unnecessarily
small timesteps, since the Solar gravity is much larger than the forces
from neighbour particles.

Our scheme is summarized as follows:
\begin{enumerate}
 \item Make a tree at time $t$ and calculate accelerations due to
       $H_{Soft}$.
 \item Give a velocity kick.
 \item Integrate positions and velocities from $t$ to $t+\Delta t$ using
       the Hermite scheme with the block timestep and $H_{Hard}$.
 \item Go back to step 1.
\end{enumerate}
When we integrate $H_{Hard}$, we use the list of neighbours for
particles to save the calculation time. If we do not use the neighbour
lists, we have to calculate forces from all particles and the
calculation cost becomes $O(N^2)$. We construct the neighbour list of
particle $i$ by selecting particles within distance $r_{nl}$ from
particle $i$ at time $t$. Here $r_{nl}$ must be sufficiently larger than
$r_{cut}$ , so that particles outside the radius $r_{nl}$ do not enter
the sphere of radius $r_{cut}$ during one timestep. In this paper, we
use $r_{nl} \gtrsim r_{cut} + 3 \Delta t \sigma$, where $\sigma$ is the
the velocity dispersion. In order to find neighbours fast, we use a
uniform 2-D grid with the grid size smaller than $r_{cut}$. To summarize
our neighbour finding way, we first assign all particles to cells. We
then look over the neighbouring cells which are within $r_{nl}$ from a
particle. The particles in the neighbouring cells are its neighbours.

 \section{Accuracy and Performance} 

In this section, we present the result of test calculations for the
accuracy and performance of our new algorithm.  We adopted the
distribution of planetesimals following the Hayashi model for test
calculations. The surface density at 1 AU is 10 $\rm g/cm^2$. The unit
mass, the gravitational constant $G$, and the unit length are normalized
to one solar mass, one, and 1 AU respectively. For most of the tests, we
use 10000 equal-mass particles distributed randomly between the radii of
0.9 and 1.0 AU. Their mass is $1.45\times 10^{23}$ g and their
velocities follow the Rayleigh distribution with $<e>=5r_H, <i>=2.5r_H$,
where $<e>$ and $<i>$ are the dispersions of the eccentricity and
inclination. Initial radius of particles is 364 km, which corresponds to
the density of 3 $\rm g/cm^3$. Physical collisions are handled under the
assumption of perfect accretion. The radius of the collision product is
determined to keep the density unchanged. Unless specified otherwise,
all test calculations are for 10 orbital periods.

\begin{figure}
  \begin{center}
   \FigureFile(160mm,80mm){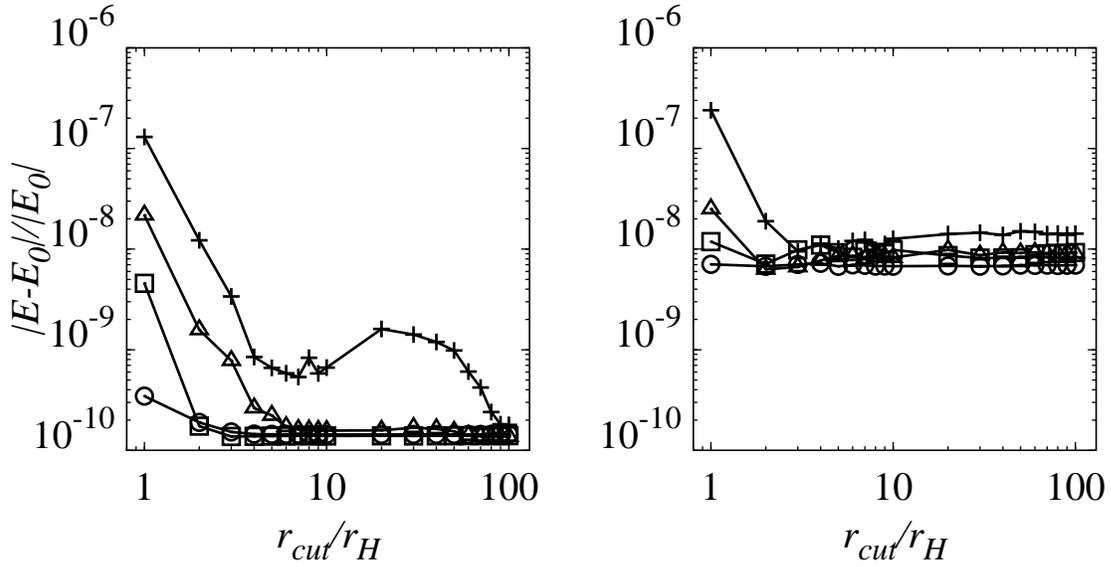}
  \end{center}
  \caption{The energy error plotted against the cutoff radius. Crosses,
  triangles, squares and circles show the results with $\Delta t=0.04,
  0.02, 0.01$ and $0.005$ yr, respectively. The left and right panels
  show the results with $\theta=0.1$ and 0.5,
  respectively.}\label{fig:energy_splrc}
 \end{figure}

\begin{figure}
  \begin{center}
   \FigureFile(80mm,80mm){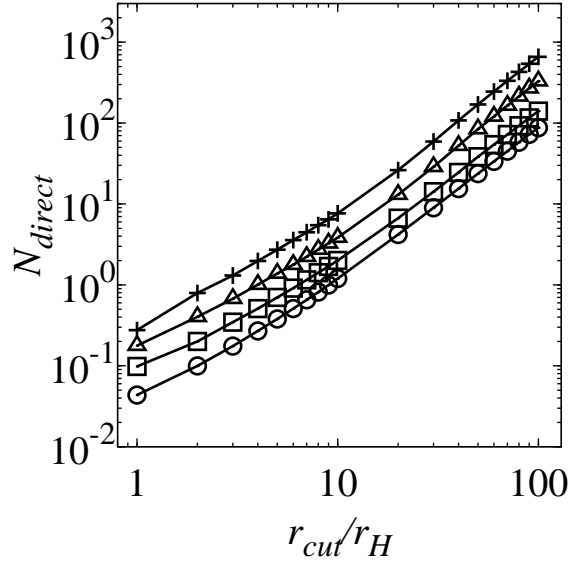}
  \end{center}
  \caption{The number of direct gravitational interactions per one tree
  timestep per one particle with the spline function plotted against the
  cutoff radius. Crosses, triangles, squares and circles show the
  results with $\theta=0.1, \Delta t=0.04, 0.02, 0.01$ and $0.005$ yr,
  respectively.}\label{fig:ndirect_splrc}
 \end{figure}

\begin{figure}
  \begin{center}
   \FigureFile(80mm,80mm){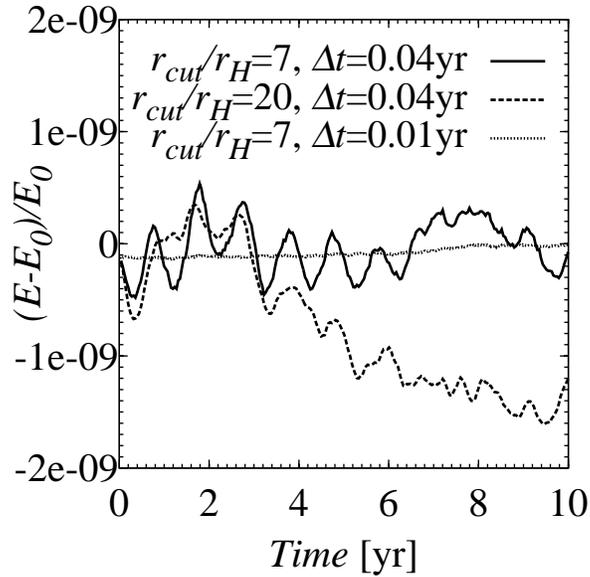}
  \end{center}
  \caption{The relative energy error of the system with the spline
  function plotted against the calculation time. The solid curve shows
  the result with $\Delta t=0.04$ yr, $r_{cut}/r_H=7$ and $\theta=0.1$. The
  dashed curve shows the result with $\Delta t=0.04$ yr, $r_{cut}/r_H=20$
  and $\theta=0.1$. The dotted curve shows the result with $\Delta
  t=0.01$ yr, $r_{cut}/r_H=7$ and
  $\theta=0.1$.}\label{fig:time_energy_splrc}
 \end{figure}

Figures \ref{fig:energy_splrc} and \ref{fig:ndirect_splrc} show the
relative energy error of the system, $|E-E_0|/|E_0|$ where $E_0$ is the
energy of the system at time 0, and the number of direct gravitational
interactions per one tree timestep per one particle, $N_{direct}$, as a
function of the cutoff radius $r_{cut}$, for the case of the spline
changeover function (eq. \ref{eq:spline}). The cutoff radius $r_{cut}$
is normalized by the Hill radius $r_H$ at 1 AU. In figure
\ref{fig:energy_splrc}, we plot the largest energy error during the time
integration for 10yr (10 periods). We use $\eta=0.05$,
timestep $\Delta t=0.040-0.0050$ yr, and opening angle $\theta=0.5$ and
$0.1$. The relative energy error is practically independent of $r_{cut}$
if $r_{cut}/r_H> 3$ and $\Delta t < 0.020$ yr. The number of mutual interactions is
proportional to the square of the cutoff radius (figure
\ref{fig:ndirect_splrc}). Since the scale height
of the disk is about $5r_H$, it is in most cases smaller than
$r_{cut}$. Therefore the number of particles is proportional to the
square of radius. Figures \ref{fig:energy_splrc} and
\ref{fig:ndirect_splrc} show that by using $r_{cut}/r_H \sim 3$, we can
achieve high accuracy with very small value of $N_{direct}$. On the
other hand, the increase in the calculation cost is pretty small even
when we use very large $r_{cut}/r_H$ such as $50$, because the force
calculation using the tree is more expensive. In the case of $\theta =
0.5$, the energy error is lower-bounded at $10^{-8}$, while with $\theta
= 0.1$, the error can be reduced to $10^{-10}$. In this case, using
$\Delta t=0.040$ yr resulted in a rather large error as shown in figure
\ref{fig:energy_splrc}. This behaviour can be understood by looking at
the time evolution of the error. Figure \ref{fig:time_energy_splrc}
shows the time evolution of the error for $\Delta t=0.040$ yr and
$0.010$ yr. The behaviour of the error shows quasi-periodic behaviour
with a period $\sim 1$ yr for $\Delta t=0.040$ yr. In figure
\ref{fig:energy_splrc}, the relative energy error for $\Delta t=0.040$
yr, $r_{cut}/r_H=20$ is larger than that for $\Delta t=0.040$ yr,
$r_{cut}/r_H=7$. This is because the random energy error from tree scheme
with large timestep is dominant for $\Delta t=0.040$ yr. Thus, this
peculiar behaviour occurred.

\begin{figure}
  \begin{center}
   \FigureFile(160mm,80mm){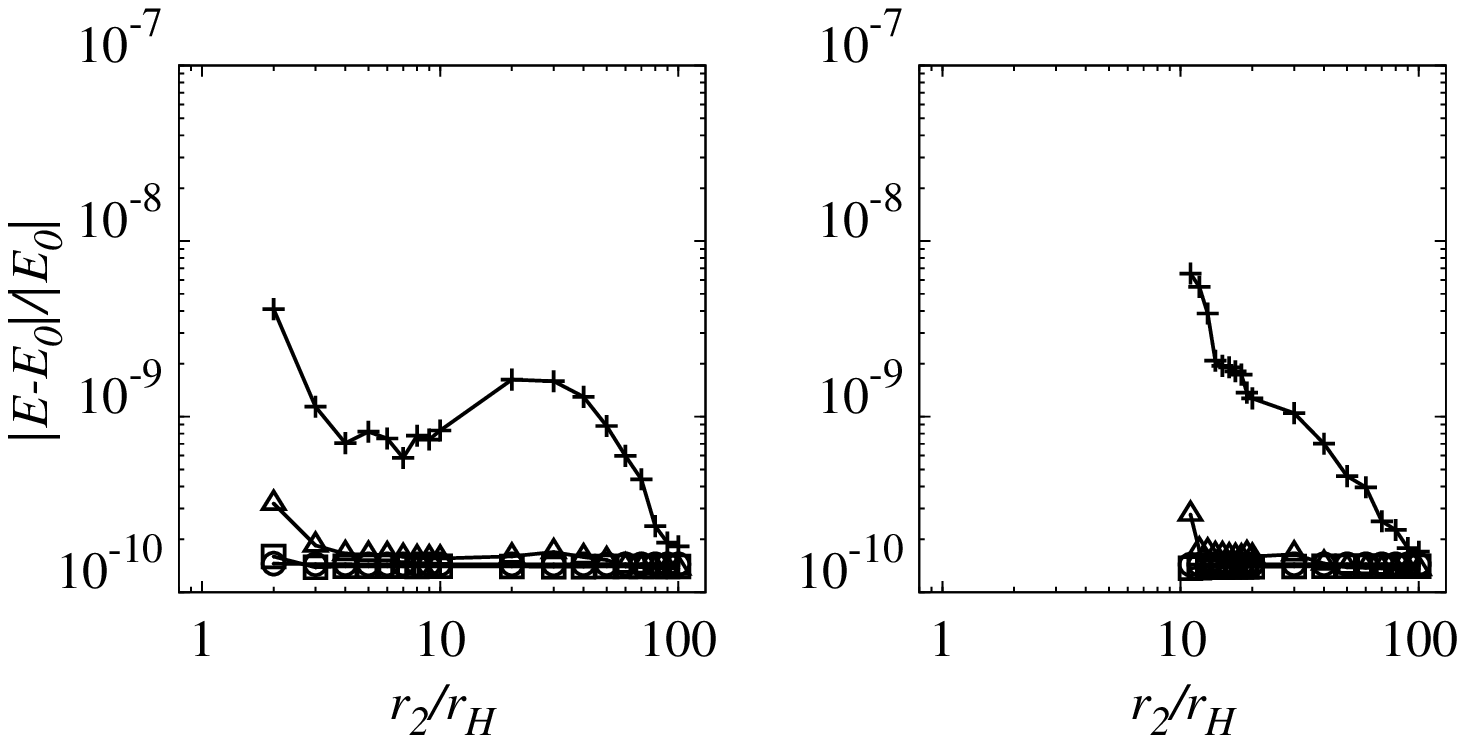}
  \end{center}
  \caption{The relative energy error of the system with the DLL function
  plotted against the $r_2$ cutoff radius. Crosses, triangles, squares
  and circles show the results with $\Delta t=0.04, 0.02, 0.01$ and
  $0.005$ yr, respectively. The left and right panels show the results
  with $r_1/r_H=1, r_2/r_H=2-100, \theta=0.1$ and $10, 11-100, 0.1$,
  respectively.}\label{fig:energy_dllrc_th01}
 \end{figure}

\begin{figure}
  \begin{center}
   \FigureFile(160mm,80mm){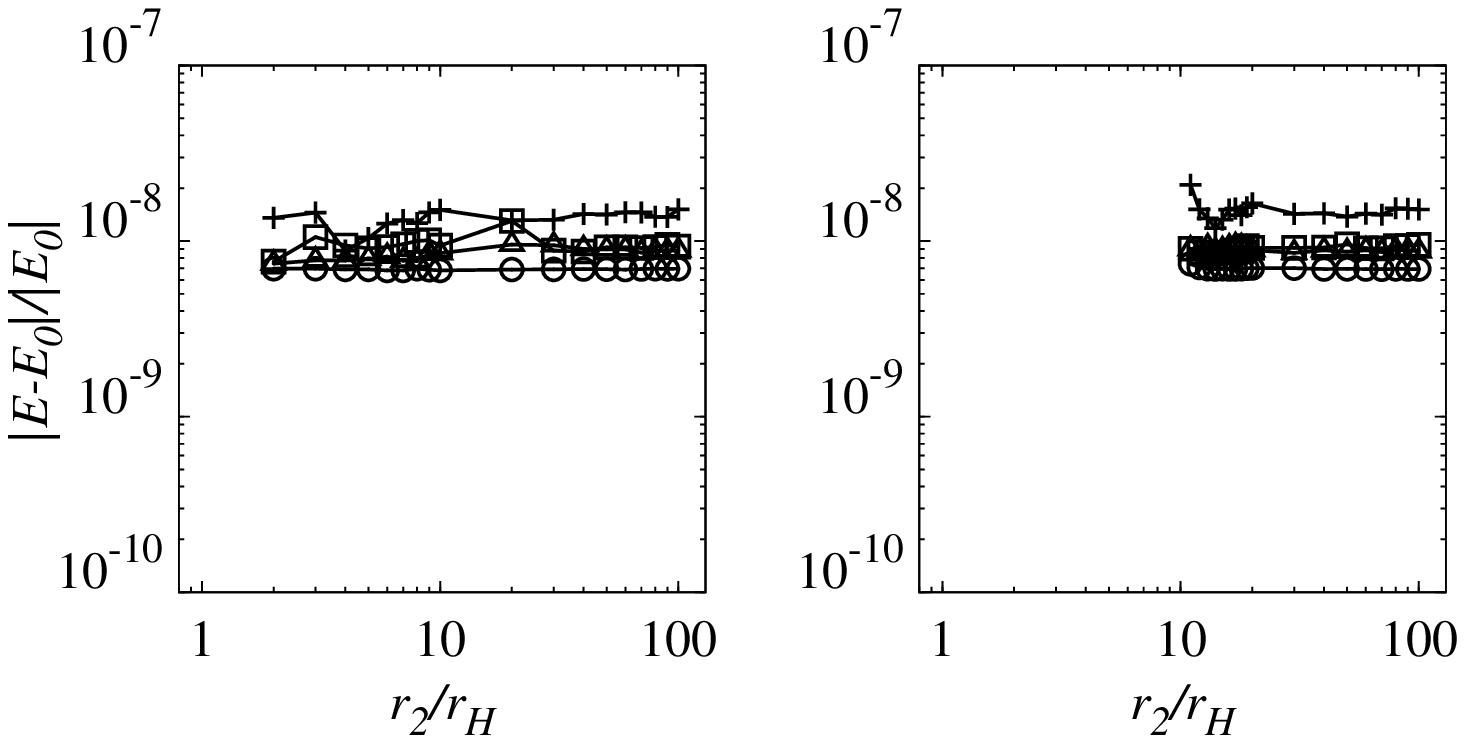}
  \end{center}
  \caption{The relative energy error of the system with the DLL function
  plotted against the $r_2$ cutoff radius. Crosses, triangles, squares
  and circles show the results with $\Delta t=0.04, 0.02, 0.01$ and
  $0.005$ yr, respectively. The left and right panels show the results
  with $r_1/r_H=1, r_2/r_H=2-100, \theta=0.5$ and $10, 11-100, 0.5$,
  respectively.}\label{fig:energy_dllrc_th05}
 \end{figure}

Figures \ref{fig:energy_dllrc_th01} and \ref{fig:energy_dllrc_th05} show
the energy error for the case of the DLL cutoff functions, with the inner
radius $r_1/r_H=1$ and $10$. We can see that the behaviour of the error
is quite similar to that in the case of the spline cutoff, and
independent of the choice of $r_1$.

\begin{figure}
  \begin{center}
   \FigureFile(80mm,80mm){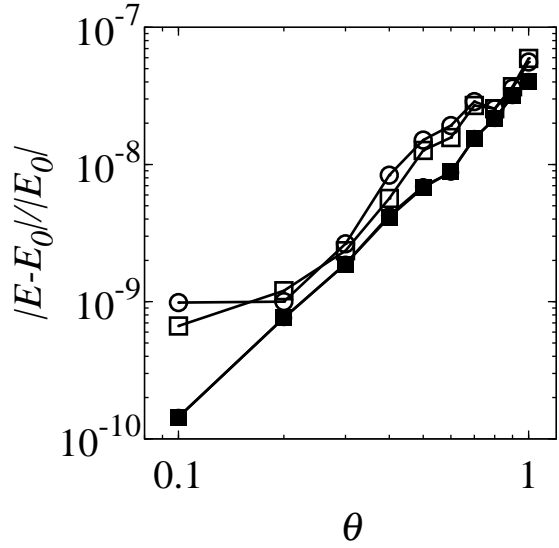}
  \end{center}
  \caption{The relative energy error of the system with the spline
  function plotted against the opening angle $\theta$. Open squares,
  open circles, filled squares and filled circles show the results with
  $r_{cut}/r_H=10$ and $\Delta t=0.04$ yr, $r_{cut}/r_H=50$ and $\Delta t=0.04$
  yr, $r_{cut}/r_H=10$ and $\Delta t=0.005$ yr and $r_{cut}/r_H=50$ and $\Delta
  t=0.005$ yr, respectively.}\label{fig:energy_splth}
 \end{figure}

\begin{figure}
  \begin{center}
   \FigureFile(80mm,80mm){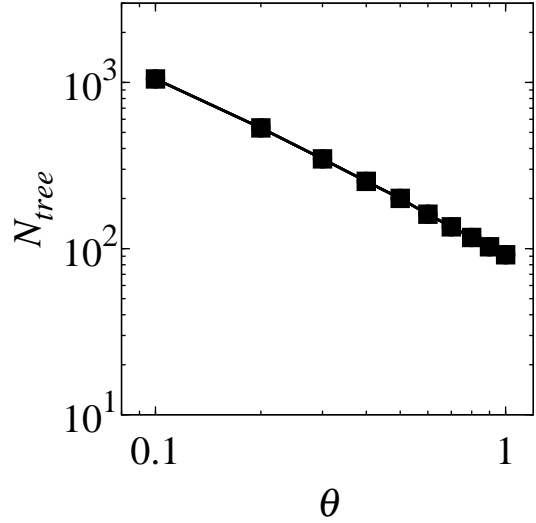}
  \end{center}
  \caption{The number of tree gravitational interactions per one tree
  timestep per one particle with the spline function plotted against
  the opening angle $\theta$. Open squares, open circles, filled squares
  and filled circles show the results with $r_{cut}/r_H=10$ and $\Delta
  t=0.04$ yr, $r_{cut}/r_H=50$ and $\Delta t=0.04$ yr, $r_{cut}/r_H=10$
  and $\Delta t=0.005$ yr and $r_{cut}/r_H=50$ and $\Delta t=0.005$ yr,
  respectively. The four results are practically
  indistinguishable.}\label{fig:ntree_splth}
 \end{figure}

\begin{figure}
  \begin{center}
   \FigureFile(160mm,80mm){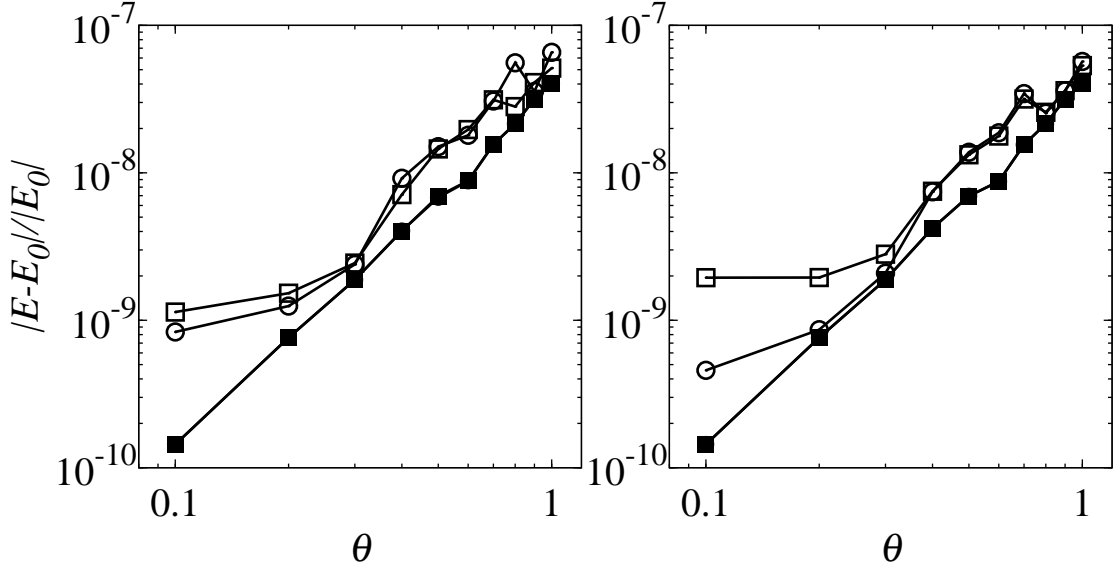}
  \end{center}
  \caption{The relative energy error of the system with the DLL
  function plotted against the opening angle $\theta$. The left panel
  shows the results with $r_1/r_H=1$ and open squares, open circles,
  filled squares and filled circles show the results with $r_2/r_H=3$
  and $\Delta t=0.04$ yr, $r_2/r_H=10$ and $\Delta t=0.04$ yr,
  $r_2/r_H=3$ and $\Delta t=0.005$ yr and $r_2/r_H=10$ and $\Delta
  t=0.005$ yr, respectively. The right panel shows the results with
  $r_1/r_H=10$ and open squares, open circles, filled squares and filled
  circles show the results with $r_2/r_H=15$ and $\Delta t=0.04$ yr,
  $r_2/r_H=50$ and $\Delta t=0.04$ yr, $r_2/r_H=15$ and $\Delta t=0.005$
  yr and $r_2/r_H=50$ and $\Delta t=0.005$ yr,
  respectively.}\label{fig:energy_dllth}
 \end{figure}

Figures \ref{fig:energy_splth} and \ref{fig:ntree_splth} show the
relative energy error and the number of interactions per one particle in
the tree part as a function of the opening angle $\theta$. In figure
\ref{fig:energy_splth}, the energy error shows the power-law dependence
as $\propto \theta^{2.5}$ for the case of $\Delta t=0.0050$ yr. On the
other hand, for $\Delta t=0.040$ yr, the error dose not go below
$10^{-9}$ for small values of $\theta$. This is because, for $\Delta
t=0.040$ yr, the truncation error of the integrator becomes larger than
the error due to force approximation for $\theta = 0.1$. In figure
\ref{fig:ntree_splth}, the number of interactions per one particle is
about two orders of magnitudes smaller than that for direct calculation,
for $\theta=0.1$. In figure \ref{fig:ntree_splth}, we can see that the
dependence of $N_{int}$ to $\theta$ is rather weak. In the case of
stellar systems, calculation cost is proportional to $\theta^{-2 \sim
-3}$ (\cite{m91}). In our experiment the dependence is $\sim
\theta^{-1}$.  This is because the distribution of particles is a thin
and narrow ring.  Figure \ref{fig:energy_dllth} shows the relative
energy error as a function of opening angle for the case of the DLL
function again, the behaviour of the error is similar to the case of
spline function.

\begin{figure}
   \begin{center}
    \FigureFile(80mm,80mm){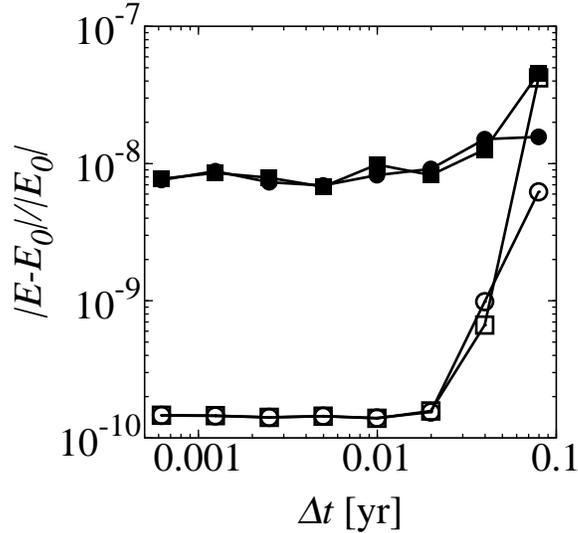}
   \end{center}
   \caption{The relative energy error of the system with the spline
  function plotted against the tree timestep. Open squares, open
  circles, filled squares and filled circles show the results with
  $\theta=0.1$ and $r_{cut}/r_H=10$, $\theta=0.1$ and $r_{cut}/r_H=50$,
  $\theta=0.5$ and $r_{cut}/r_H=10$, $\theta=0.5$ and $r_{cut}/r_H=50$,
  respectively.}\label{fig:energy_spldt}
  \end{figure}

Figure \ref{fig:energy_spldt} shows the relative energy error as a
function of the size of timestep for $H_{Soft}$. In the case of
$\theta=0.1$ and $\Delta t<0.010$ yr, the error is dominated by that
from the tree approximation, and becomes independent of $\Delta t$. If
we want to keep the error in 10yr ($10$ periods) to be less than
$10^{-9}$, a pair of $\theta \sim 0.2$ and $\Delta t \sim 0.02$ is
probably a good choice. In realistic calculations $10^{-9}$ in 10 orbits
is probably okay, though how small the error should be is a difficult
question. At least, the error of order of $10^{-9}$ is smaller than that
of energy change of planetesimals due to gas drag and collisional
damping. We do not show the result for the DLL cutoff here. We
calculated by using same parameters for the DLL cutoff and confirmed
that the result is essentially the same as that for the spline cutoff in
figure \ref{fig:energy_spldt}.

\begin{figure}
  \begin{center}
   \FigureFile(80mm,80mm){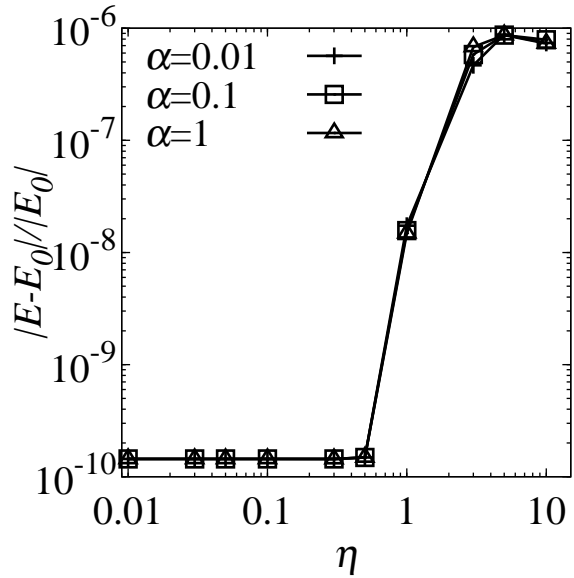}
  \end{center}
  \caption{The relative energy error of the system plotted against the
  timestep accuracy parameter $\eta$. Crosses, squares and triangles
  show the results with $\alpha=1, 0.1$ and $0.01$,
  respectively.}\label{fig:energy_epseta}
 \end{figure}

\begin{figure}
  \begin{center}
   \FigureFile(80mm,80mm){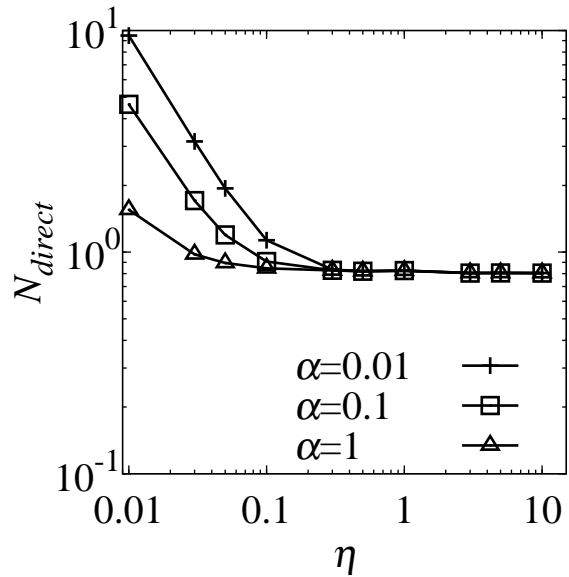}
  \end{center}
  \caption{The number of direct gravitational interactions per one tree
  timestep per one particle with the spline function plotted against the
  timestep accuracy parameter $\eta$. Crosses, squares and triangles
  show the results with $\alpha=1, 0.1$ and $0.01$,
  respectively.}\label{fig:ndirect_epseta}
 \end{figure}

Figures \ref{fig:energy_epseta} and \ref{fig:ndirect_epseta} show the
relative energy error and the calculation cost of neighbour force
$N_{direct}$ as functions of the accuracy parameter $\eta$. Other
parameters are $\theta=0.1, r_{cut}/r_H=10$ and $\Delta t=0.0050$
yr. Figure \ref{fig:energy_epseta} shows that the energy error is
practically independent of the choice of $\alpha$. On the other hand, in
figure \ref{fig:ndirect_epseta}, small $\alpha$ results in the increase
of $N_{direct}$. In practice, a pair of $\eta = 0.1$ and $\alpha=1$
seems to be a good choice.

\begin{figure}
  \begin{center}
   \FigureFile(80mm,80mm){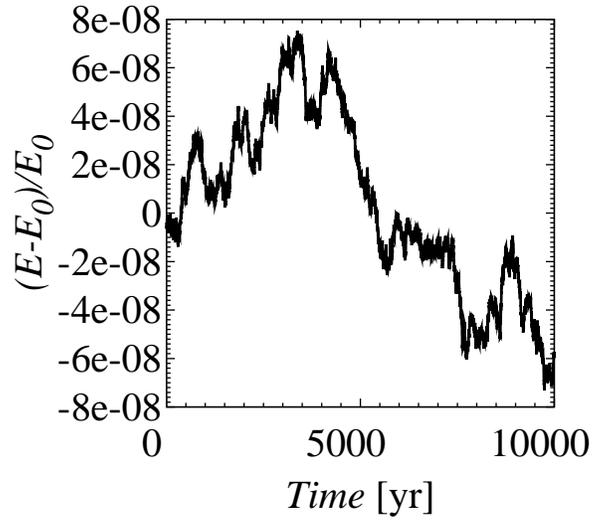}
  \end{center}
  \caption{Energy error of the system plotted against a function of
  time.}\label{fig:energy_long}
 \end{figure}

Figure \ref{fig:energy_long} shows the long-term variation of the
relative energy error. The calculation is done with the opening angle
$\theta=0.5$, the cutoff radius $r_{cut}/r_H=10$ and the timestep
$\Delta t= 0.0050$ yr. In this case, the energy error reaches about $7.5
\times 10^{-8}$ after the time integration for $10^4$ yr ($10^4$ orbital
periods), while it is $9.8 \times 10^{-9}$ for 10 yr ($10$ periods). In
other words, the energy error grows 10 times larger as the integration
time becomes 1000 times longer.  It shows that the growth of energy
error is stochastic like a random walk. It means that the error is
mainly caused by the force error of the tree scheme (\cite{bh89}). The
growth of energy error is, therefore, expected to be slow and the error
is small enough even after long calculations.

\begin{figure}
  \begin{center}
   \FigureFile(80mm,80mm){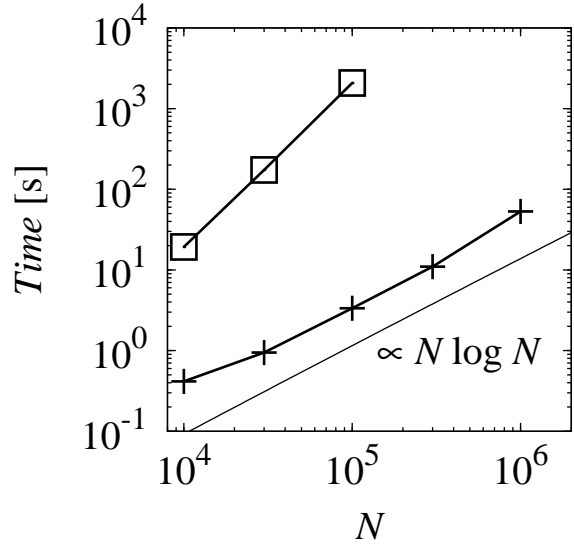}
  \end{center}
  \caption{Calculation time per 0.0050 yr plotted against a function of
  number of particles. Crosses and squares show the results of PPPT and
  fourth-order Hermite scheme, respectively.}\label{fig:time}
 \end{figure}

Figure \ref{fig:time} shows the calculation time per one tree timestep
as a function of the total number of particles in the system $N$. The
calculation is done with the opening angle $\theta=1$, the cutoff
radius $r_{cut}/r_H=5$ and the timestep $\Delta t=0.0050$ yr. We
used the Intel(R) Core(TM)2 Quad CPU Q6600(2.4GHz). It shows that the
calculation time increases as $O(N \log N)$. Therefore, we reduce the
calculation cost from $O(N^2)$ to $O(N \log N)$.

\begin{figure}
  \begin{center}
   \FigureFile(80mm,80mm){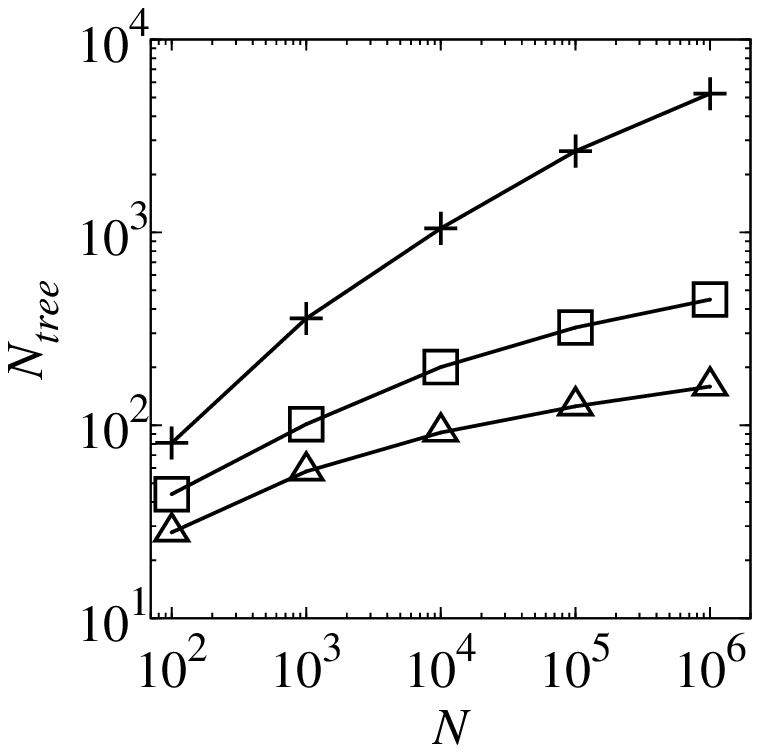}
  \end{center}
  \caption{The number of tree gravitational interactions per one tree
  timestep per one particle with the spline function plotted against
  a function of number of particles. Crosses, open squares and open
  triangles show the results with $\theta=0.1, 0.5$ and 1,
  respectively.}\label{fig:tree_number}
 \end{figure}

Figure \ref{fig:tree_number} shows the number of tree interactions,
$N_{tree}$ per particle as a function of $N$. We can see that $N_{tree}$
is roughly proportional to $O(\log N)$.

 \section{Summary and Discussion} 

We have developed a new hybrid $N$-body simulation algorithm for the
simulation of collisional $N$-body systems. This new scheme is
constructed by combining the tree and direct schemes using the hybrid
integrator. The results of test simulations of evolution of a
planetesimal system show that our new scheme PPPT can drastically reduce
the calculation cost, to the level comparable to the cost of a tree
scheme with constant timestep while keeping accuracy sufficient for
realistic simulations.

In principle, our scheme can be used for collisional systems other than
planetary systems, such as globular clusters or stars around a
supermassive blackhole in the galactic center. We'll show the results of
simulations of such systems using our scheme in future.

\bigskip

We are grateful to Tomoaki Ishiyama, Masaki Iwasawa, Michiko Fujii,
Kuniaki Koike, Keigo Nitadori and Yusuke Tsukamoto for fruitful
discussions. This work was partially supported by the Research Fund
for Students (2009) of the Department of Astronomical Science, the
Graduate University for Advanced Studies, KAKENHI 21244020 and 21220001.


\begin{thebibliography}{}
 \bibitem[Aarseth(1963)]{a63}
		 Aarseth, S. J. 1963, \mnras, 126, 223
 \bibitem[Aarseth(2003)]{a03} 
		 Aarseth, S. J. 2003, Gravitational N-Body
		 Simulations: Tools and Algorithms (Cambridge Monographs
		 on Mathematical Physics)
 \bibitem[Abe et al.(1986)]{asio86}
		 Abe, H., Sakairi, N., Itatani, R., Okuda, H., 1986,
		 Comput. Phys., 63, 247
 \bibitem[Bagla(2002)]{b02}
		 Bagla, J. S. 2002, JA\&A, 23, 185
 \bibitem[Barnes \& Hut(1986)]{bh86}
		 Barnes, J., \& Hut, P. 1986, \nat, 324, 446
\bibitem[Barnes \& Hut(1989)]{bh89}
		 Barnes, J., \& Hut, P. 1989, \apjs, 324, 446
 \bibitem[Brunini \& Viturro(2003)]{bv03}
		 Brunini, A., \& Viturro, H. R. 2003, \mnras, 346, 924
 \bibitem[Brunini et al.(2007)]{bsvc07}
		 Brunini, A., Santamar\'{i}a, P. J., Viturro, H. R., \& Cionco,
		 R. G. 2007, P\&SS, 55, 2121
 \bibitem[Bulirsch \& Stoer(1964)]{bs64}
		 Bulirsch, R., Stoer, J. 1964, Num. Math. 6, 413
 \bibitem[Calvo \& Sanz-Serna(1993)]{cs93}
		 Calvo, M. P., \& Sanz-Serna, J. M. 1993, SIAM
		 J. Sci. Comput., 14, 4, 936
 \bibitem[Chambers(1999)]{c99}
		 Chambers, J. E.\ 1999, \mnras, 304, 793
 \bibitem[Dubinski et al.(2004)]{dkph04}
		 Dubinski, J., Kim, J., Park, C., \& Humble, R. 2004, New Astron., 9, 111
 \bibitem[Duncan, Levison, \& Lee(1998)]{dll98}
		 Duncan, M. J., Levison, H. F., \& Lee, M. H. 1998, \aj, 116, 2067
 \bibitem[Fehlberg(1968)]{f68}
		 Fehlberg, E. 1968, NASA TR R 287
 \bibitem[Fujii et al.(2007)]{fifm07}
		 Fujii, M., Iwasawa, M., Funato, Y., \& Makino, J. 2007, \pasj, 59,
		 1095
 \bibitem[Hockney \& Eastwood (1981)]{he81}
		 Hockney, R.W., \& Eastwood, J.W. 1981, Computer Simulation Using
		 Particles (New York: McGraw-Hill)
 \bibitem[Ishiyama et al.(2009)]{ifm09}
		 Ishiyama, T., Fukushige, T., \& Makino, J. 2009, \pasj, 61, 1319
 \bibitem[Kinoshita, Yoshida \& Nakai(1991)]{kyn91}
		 Kinoshita, H., Yoshida, H., \& Nakai, H. 1991,
		 Celest. Mech. Dyn. Astron., 50, 59
 \bibitem[Levison \& Duncan(2000)]{ld00}
		 Levison, H. F., \& Duncan, M. J. 2000, \aj, 120, 2117
 \bibitem[Makino \& Aarseth(1992)]{ma92}
		 Makino, J., \& Aarseth, S. J. 1992, \pasj, 44, 141
 \bibitem[Makino(1991)]{m91}
		 Makino, J. 1991, \pasj, 43, 859
 \bibitem[McMillan \& Aarseth(1993)]{ma93}
		 McMillan, S. L. W., \& Aarseth, S. J. 1993, \apj, 414,
		 200
 \bibitem[Moore, Quillen \& Edgar(2008)]{mqe08}
		 Moore, A. J., Quillen, A. C., \& Edgar, R. G. 2008, arXiv0809.2855
 \bibitem[Skeel \& Biesiadecki(1994)]{sb94}
		 Skeel, R. D. \& Biesiadecki, J. J. 1994,
		 Ann. Numer. Math, 1, 191
 \bibitem[Skeel \& Gear(1992)]{sg92}
		 Skeel, R. D. \& Gear, C. F. 1992, Physica, D60, 311
 \bibitem[Springel(2005)]{s05}
		 Springel, V. 2005, \mnras, 364, 1105
 \bibitem[Stoer \& Bulirsch(1980)]{sb80}
		 Stoer, J., \& Bulirsch, R. 1980, Introduction to Numerical
		 Analysis (Springer Verlag, New York)
 \bibitem[Wisdom \& Holman(1991)]{wh91}
		 Wisdom, J., \& Holman, M. 1991, \aj, 102, 1528
 \bibitem[Xu(1995)]{x95}
		 Xu, G. 1995, \apjs, 98, 355
 \bibitem[Yoshikawa \& Fukushige(2005)]{yf05}
		 Yoshikawa, K., \& Fukushige, T. 2005, \pasj, 57, 849
\end{thebibliography}
\end{document}